# IDENTIFICATION OF TEST STRUCTURES FOR REDUCED ORDER MODELING OF THE SQUEEZE FILM DAMPING IN MEMS

*Aurelio Somˆ, Giorgio De Pasquale*

Laboratory of Microsystems, Department of Mechanics, Politecnico di Torino,
C.so Duca degli Abruzzi, 24 - 10129 Torino, Italy.
aurelio.soma@polito.it, giorgio.depasquale@polito.it

## ABSTRACT

In this study the dynamic behaviour of perforated microplates oscillating under the effect of squeeze film damping is analyzed. A numerical approach is adopted to predict the effects of damping and stiffness transferred from the surrounding ambient air to oscillating structures; the effect of hole's cross section and plate's extension is observed. Results obtained by F.E.M. models are compared with experimental measurements performed by an optical interferometric microscope.

## 1. INTRODUCTION

Micro Electro Mechanical Systems (MEMS), characterized by many different features and shapes, are currently realized by numerous oscillating elements and components; may be difficult to recognize common parts among a so widely diffused devices population but we observe that oscillating parts, which shape is derived from the classical beam or plate, is largely widespread. To investigate interactions of suspended plates with surrounding gas during their oscillation is then important to predict their operating behaviour; these plates are usually the upper part of a variable capacitor. Dynamic is not strongly influenced by inertial force because of the smallness of structural mass, but is strictly dependent from damping and stiffness forces transferred from the air present in the thin gap separating the movable structure and substrate; the order of magnitude of this contributions, identified by the name of *squeeze film damping*, are frequently higher than the internal damping and stiffness ones.

### 1.1. Literature review

Many investigations have been recently realized on the squeeze film damping effect; Veijola *et al* [1] proposed a compact model to consider the border flow at structure edges by extending surface dimensions and producing an equivalent damping effect. The model is referred to micro-dampers with gap size comparable to surface dimensions. Main limitation is the assumption of an incompressible flow that restricts analysis validity to not too high frequency values. Hutcherson and Wenjing [2] examined mechanical resonators behaviour at very low pressures and in free-molecule regime conditions; they developed a dynamic simulation code for quality factor prediction, which results are in good agreement with experimental data. They observed that quality factor depends linearly on frequency and strongly depends on the amplitude when it is large (comparable to the gap) and is independent from amplitude when it is small. The model only considers elastic collisions between molecules and structure.

Veijola [3] studied the effect of gas inertia in a squeezed-film damper both analytically and numerically, observing that this property can not be neglected above the cut-off frequency. He derived equivalent-circuit mechanical impedance and admittance to solve linearized Reynolds equation in continuum flow, with a small velocity and by neglecting border effects. Another analytical model [4] is derived to estimate squeeze film damping under perforated plates based on the spatial repetition of a single cell; it takes into account both the gas flow in the air gap and in the perforations. Rarefaction effect in the slip flow regime is considered and results are checked by F.E.M. simulations. A repetitive pattern of the pressure distribution around each hole in perforated plates is also observed by Mohite *et al* [9] that identified independent pressure cells of circular geometry, analytically solved by a one-dimensional Reynolds equation in polar coordinates. The complex pressure obtained is used to identify the stiffness and damping offered by the pressure cell, and then added up separately to extract global dynamic parameters.

Air damping acting on torsion mirrors devices is the goal of some important studies; Bao *et al* [5] evaluated the squeeze-film air damping of a rectangular torsion mirror at finite normalized tilting angles by an analytical model derived from a nonlinear Reynolds equation; damping pressure and damping torque are derived as functions of the tilting angle and the aspect ratio. Results show that damping torque is highly nonlinear dependent from tilting angle and basically linear dependant from the aspect ratio of the mirror. Minikes *et al* [15] analyzed damping and quality factors in resonating torsion mirrors in presence of both ambient and low pressures; two independent experimental damping extraction methods produced results in good agreement with existing theoretical models, also if some inaccuracies are observed at vacuum condition.

Chang *et al* [18] analyzed the dynamical behaviour of a torsion mirror under the effect of squeeze film air damping, gas rarefaction and surface roughness. Pan *et al* [19] studied the dynamic behaviour of a torsion mirror by deriving an analytical solution for the effect of squeeze film damping from the linearized Reynolds equation under the assumption of small displacements. Damping pressure and torque are expressed in function of the rotational angle and the angular velocity of the mirror. Numerical simulations and experimental measurements confirmed the validity of analytical model.

Pursula *et al* [6] simulated the behaviour of a planar gas-damped accelerometer under electrostatic loading by finite element method; the transient model combines electro-mechanical coupling to nonlinear squeeze-film damping effects and utilize various reduced-order and reduced-dimensional methods to significantly reduce the computational cost of the simulation. Nayfeh and Younis [8] presented a new approach to the modeling and simulation of flexible microstructures under the effect of squeeze-film damping by expressing the pressure distribution in terms of the structural mode shapes. The resulting governing equation is then solved using a F.E.M. to extract natural frequencies, structural mode shapes, corresponding pressure distribution and quality factor. Zhang *et al* [10] analyzed the effect of squeeze film damping by the coupling of elastic beam theory and Reynolds equation for isothermal incompressible gas films; they obtained an expression of quality factor and resonant frequencies in function of two dimensionless parameters depending on physical properties of the beam and gas. Pandey and Pratap [11] studied the effect of squeeze film damping under the effect of gas rarefaction and surfaces roughness; they observed that rarefaction lowers damping force without affect spring force significantly, but in presence of a high surface roughness and amplitude vibration, both damping and spring forces increases considerably.

Steeneken *et al* [12] studied the dynamic behaviour of a capacitive RF shunt switch by extracting its motion from a capacitance measurement at high time resolution and comparing results with analytical and numerical ones. An other comparison between squeeze film damping effects on an inertial sensor at high and low pressure of surrounding fluid is performed by Braghin *et al* [13], that used a multi-physic code for numerical simulations and experimental measurements.

Lee *et al* [14] investigated the size effect on the quality factor associated with the first mode of microcantilever vibration in ambient air, by the comparison between numerical models, analytical analysis and experimental measurements; existing approximate models for air damping evaluation are modified to account geometry effects, demonstrating that quality factors are proportional to geometrical dimensions. The introduction of an 'effective damping width' of the structure allows to Bao *et al* [16] the easy treatment of boundary effect on damping; they found an analytical expression of damping pressure for rectangular plates with holes by the modification of Reynolds equation with a term related to the damping effect of gas flow through holes. An other investigation [17] lead to a model for the estimation of microstructures air damping in low vacuum by using an energy transfer mechanism instead the momentum transfer mechanism in Christian's model [22], taking into account nearby objects (e.g. the electrodes for electrostatic driving), the dimensions of vibrating plates and the air gap extension.

In a previous work, finally, Somà *et al* [7] studied the effect of holes dimensions on dynamic behaviour of a suspended oscillating plate; damping and stiffness contribution of gas film are numerically derived for a wide frequency range and for different holes dimensions.

In this paper the non-linear coefficients of equivalent damping and stiffness is studied by using finite element models. Two numerical models have been used: the first one considers only the fluidic domain, assuming that MEMS device is a rigid body; the second one couples the fluidic domain to that structural too, considering therefore the flexibility of the device too. Finally test structure is designed in order to enhance the effect of some structural and process parameters such as holes and geometrical parameters. The experimental results is obtained with a laser interferometer measurement technique in terms of

Frequency Response Function (FRF)

## 2. TEST STRUCTURES

Experimental measurements are conducted on test structures specifically designed for this study, realized by polysilicon according to the microfabrication design and process rules of STMicroelectronics MEMS Business Unit (Cornaredo, Italy). They are realized in according to the so-called 'Thelma' process, consisting in a gradual growth of superposed polysilicon layers; the silicon substrate is firstly covered by a 2.5μm thick layer of permanent oxide at 1100°C, then by a polysilicon layer and by a 1.6μm thick layer of sacrificial oxide. The polysilicon suspended layer growth is then completed in the reactors, till the reaching of a 15μm thickness. After the chemical removal of sacrificial oxide layer in dry environment, a metallization is finally deposited on all the surfaces used for direct contact.

Test structures are characterized by a central suspended plate covered by square holes, connected to four lateral clamped supports of a small cross section area (figure 1); this particular shape allows an out-of-plane deflection of lateral supports during the excitation and, consequently, a quasi-rigid oscillation of the central plate, where structural stiffness is concentrated.

Different shape-types specimens are realized on the same chip, to account the effect of two geometrical parameters; in particular, they have a different plate width and cross section side of the holes. Suspended structures are 15μm thick and gap extension is 1.6μm; the polysilicon is characterized by an estimated Young modulus E=147GPa, a specific density $\rho$=2.33·10$^{-15}$kg/μm$^3$ and a Poisson ratio $\nu$=0.2152, also if physical processes and chemical reactions typical of some production phases may cause sensible variations of these nominal values.

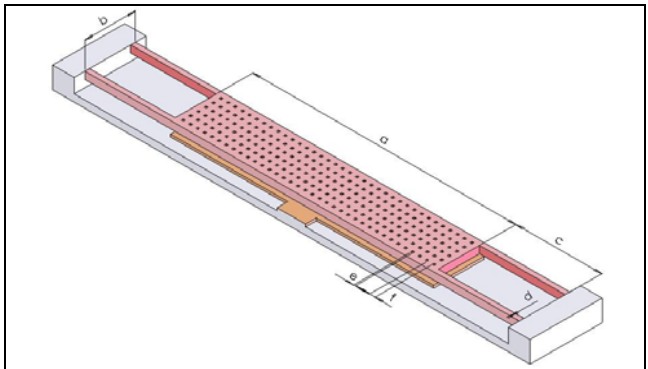

*Figure 1. Test structure type A geometry.*

Six different test structures are realized, marked with letters from A to F; first four specimens only differs by holes cross section size, that is gradually increased, E and F structures are characterized by the same holes dimension (and equal to the B-structures one) but a different plate width. We observe small differences between design dimensions and effective dimensions of specimen; this is due to the difficulty to predict exact effects of acid attacks during production processes; table 1 reports real effective geometrical dimensions of all specimens, obtained by profile measurements performed by the interferometric microscope.

| type | plate length | plate width | support length | support width | hole side | hole's inter-space |
|------|-------------|-------------|----------------|---------------|-----------|--------------------|
|      | $a$ | $b$ | $c$ | $d$ | $e$ | $f$ |
| A | 372,4 | 66,4 | 122,8 | 4,0 | 5,0 | 5,2 |
| B | 363,9 | 63,9 | 122,4 | 4,3 | 6,1 | 3,9 |
| C | 373,8 | 64,8 | 123,2 | 3,7 | 7,3 | 3,0 |
| D | 369,5 | 63,5 | 123,4 | 3,9 | 7,9 | 2,3 |
| E | 363,8 | 123,8 | 123,2 | 3,8 | 6,2 | 3,8 |
| F | 363,8 | 243,8 | 122,4 | 3,8 | 6,2 | 3,8 |

*Table 1. Effective geometrical dimensions of test structures in micron.*

Figure 2 shows an image of specimen type A; the contact pad on the left feeds suspended structure, the right one is connected to the bottom polysilicon electrode. In the figure 3 is reported a 3-dimensional reconstruction of the same structure obtained by the evaluation of z-extension from the interpretation of interferometric fringes intensity.

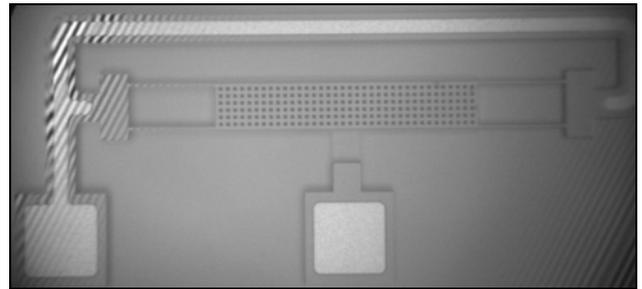

*Figure 2. Test structure type A.*

## 3. THE INTERFEROMETRIC MICROSCOPE

All experimental measurements are conducted by the optical interferometric microscope ZoomSurf 3D (Fogale Nanotech); this instrument allows surface observations of a wide material typology as metals, polymers, semiconductors, biological, etc. It is possible to perform statistical analysis of surface roughness or micromechanical devices, as well as evaluation of thin transparent layers thickness (as varnishes, plastics, glasses, etc.) of a known refraction coefficient.

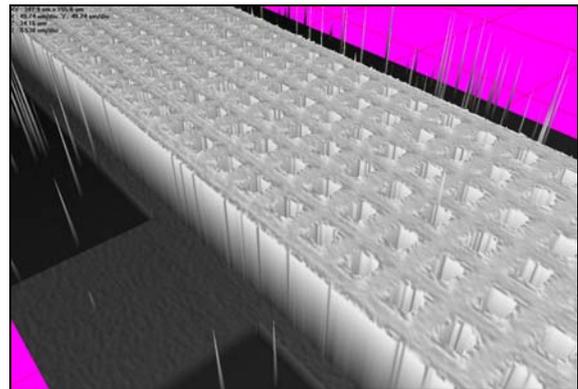

*Figure 3. 3D reconstruction of test structure type A.*

The microscope can perform profile measurements from a minimum area of 100x100μm to a maximum area of 2x2 mm. Actual setting provides a 20x objective magnification factor, that realizes a lateral resolution of 0.6μm and a vertical resolution reaching the value of 0.1nm. The maximum difference in height detectable in a single measure is 400µm. The stage is electrically moved in the plane x-y and can support a maximum weight of 5kg; the objective moves in the z-axis and can cover a distance of 250mm. Stage slope can be changed mechanically to adjust contrast fringe orientation and thickness.

Internal software allows the evaluation of quasi-static deformations of microstructures by the analysis of a series of successive measurements; it is possible to obtain a Frequency Response Function (FRF) and an amplitude oscillation map for vibrating structures under AC actuation.

The interferometric device basically is quite similar to the well-known Michelson interferometer; the light radiation is emitted from a selectable source (continuous or stroboscopic, monochromatic or polychromatic) and is then separated on two different beams. One of them is reflected by a *reference mirror*, while the other is directed on the specimen and back-reflected by its surface (or by an internal interface in the case of transparent samples); from the conjunction of these two reflected radiations the optical interference phenomena takes place, producing a new radiation to a CCD camera, which can detect the light intensity variation of each pixel separately. This variation is directly connected to the variation of the optical path of the light beam reflected by sample and can be consequently converted in the exact z-position of the corresponding reflection point of the surface. By elaborating the height data for each surface point is then possible to build a 3-dimensional reconstruction.

It is preferable to use a red monochromatic light source that generates interferometric fringes with variable light intensity and constant amplitude; measurement of fringes phase is the principle at the basis of the superficial interferometric technique called *Phase Shifting Interferometry*, the best one for rapid measurements of regular plates in motion, etching controls, micromachining processes monitoring and optical components characterization. The use of a polychromatic white light source determines, because of its intrinsic incoherent nature, a rapid decay of fringe intensity and a consequent thinner shape of them. The so-called *White Light Interferometry* technique is based on the localization of the maximum fringe contrast area and is used for superficial profile measurements of large and irregular components.

Consequently to ligth source nature variation, the software allows the modification of some optical parameters as the objective magnification factor (from 0.35x to 1.60x) and the reflection coefficient of internal reference mirror (from 5% to 85%). The software can recognize that specimen parts characterized by excessive inclination, as well as the parts covered by a layer of dust, and exclude them from the evaluation.

The microscope is equipped by a signal generator operating at both low and high frequencies up to 1MHz. Excitation voltage is provided from 0V to 200V to the structure through special conductive needles.

### 3.1. Measurement settings

Experimental tests for damping and stiffness evaluation are conducted with the *Frequency Shift* technique, consisting in the specimen actuation by a sinusoidal voltage (of a 5V amplitude and 2V offset from zero) which frequency is progressively increased by a discrete selectable step and the corresponding oscillation amplitude of the structure is registered, obtaining the FRF. Chosen actuation voltage is common for all structure types and allows the excitation far from the pull-in voltage. The first detection is performed across a wide frequency range (0-500kHz) to roughly locate fundamental resonant frequencies, reported in the subsequent table 2. Five successive identical detections are conducted in a more precise narrow frequency range centred on each resonant frequency and statistically treated, to extract the exact value of eigen-frequency and resonance amplitude [23].

This microscopic measurement method needs of a red monochromatic light source; oscillation amplitude is detected over a selectable region (active window) whose extension not corresponds to the CCD camera one, but is precisely located at the centre of suspended plate. The output oscillation amplitude value is obtained as an average of the value detected by each pixel contained in the active window. To perform a correct detection the active window must contain two interferometric fringes at least and its positioning must take into account the modal shape that will be excited (in the case of non-rigid displacements, detection errors may occur especially with an excessive active window size).

## 4. NUMERICAL SIMULATIONS

### 4.1. Frequency dependency of damping and stiffness

Many numerical simulations demonstrate that the thin fluid layer between suspended structure and substrate modifies the dynamic behaviour of oscillating element in dependence of its vibrating frequency. In the case of a low frequency harmonic excitation the air is responsible of relevant dissipative phenomena and increases global damping level of the entire system. For high frequency vibrations of movable parts the fluid become instead a source of energy accumulation, by the observation of an increase in elastic stiffness of the system, which can return outside a more relevant energy quantity, received from the environment than in the case of a vacuum ambient.

Physical considerations at the basis of this behaviour are relatively simple; when the plate moves at low velocities (that happens at low actuation frequencies), fluid viscosity is not sufficient to keep it in the same region during the entire oscillation period, that causes its motion through the holes and lateral edges, that have both a dissipative effect. If the plate velocity is higher (that corresponds at high actuation frequency values), fluid viscosity is sufficient to prevent large air particles motions, causing the molecules elastic compression. A correct plate design allows to control fluid flows through the holes and edges and to modify the global dynamic behaviour of the system; one of the most relevant design parameters is the quality factor: an high value increase the device sensitivity, but a too low damping cause excessive signals noises.

The ambient pressure modifies fluid rarefaction, that is expressed by the *Knudsen number* ($K_n$), defined as the ratio between the mean free path (corresponding to the distance covered by a molecule between two successive collisions) and a characteristic length (comparable to the gap extension). The Knudsen number relative to experimental conditions on test structures is $K_n=0.042$, that corresponds to a continuum fluid regime with the presence of slip flows causing the presence of a component of fluid velocity orthogonal to the plate motion in correspondence to the bottom surface of vibrating structure; with this fluid conditions Navier-Stokes equations are still valid, but is convenient to take into account slip flows by the introduction of an effective viscosity, defined by Veijola [20] as

$$\eta_{eff} = \frac{\eta}{1 + 9.638 K_n^{1.159}}.$$  (1)

Many numerical simulations are conducted by using commercial software (Ansys 9.0 [24]) in according to two different strategies to investigate fluidic interactions with test structures dynamics.

### 4.2. Imposed constant velocity method

By considering a plate with mass $m$ and structural stiffness $k'$ suspended on a fluid gap responsible of a viscous damping $c$ under an external electrostatic actuation $H(t)$, the corresponding governing equation of the system at low vibrating frequencies is

$$m\ddot{z} + c\dot{z} + k'z = H(t);$$  (2)

It is possible to separate the component $F(t)$ from the actuation $H(t)$, that equilibrates fluid actions only, obtaining the low frequency fluid damping expression

$$c\dot{z} = F(t),$$  (3)

that corresponds to

$$c = \frac{F(t)}{v},$$  (4)

where $v$ is the plate vertical velocity, that can be intended as an average velocity of the harmonic oscillation.

When actuation frequencies increase, fluid stiffness $k$ become relevant and the governing equation modifies as

$$m\ddot{z} + c\dot{z} + (k'+k)z = H(t);$$  (5)

by extracting the component $F(t)$ of actuation corresponding to fluid reactions, we observe that an imaginary part appears. It is natural to deduce that high frequency fluid damping is connected to the real part of this force (the only present at low frequencies) and is expressed as:

$$c\dot{z} = F_r(t),$$  (6)

where $F_r(t)$ is the real part of $F(t)$; the high frequency fluid stiffness is than connected to the imaginary part $F_i(t)$ as

$$kz = F_i(t),$$  (7)

by obtaining for a given frequency the expressions

$$c = \frac{F_r(t)}{v}$$  (8)

$$k = \frac{F_i(t) \cdot \omega}{v},$$  (9)

where is operated the substitution

$$z = \frac{v}{2\pi\omega}.$$  (10)

The complex force $F(t)$ is obtained from the complex gas pressure distribution $p$ and the bottom plate area $A$ as

$$F(t) = p(x,y,t)A.$$  (11)

From previous considerations we deduce that plate velocity and gas pressure are entities with a sinusoidal in-phase variation if actuation is at low frequencies, the appearance of a fluid compression at high frequencies instead causes a out-of-phase variation of these two parameters; here the real part of pressure distribution is still in-phase with velocity, but the imaginary one, connected to fluid stiffness, results in opposition with velocity.

Main limitation of this method is the assumption of a constant velocity value for the entire oscillating structure; particular design of specimen allows to consider a quasi-rigid motion of the central plate, but in general each structure oscillates according to a particular modal shape (or a superposition of modal shapes) for a given frequency, determining a velocity profile of the plate. In this case results obtained with the imposed velocity method must be carefully considered.

The simulation software needs to the definition of fluidic domain only; the absence of the entire structure model allows a less heavy computation, with a relevant time reduction. The fluid contained in the air gap is modelled by *fluid136* elements and the pressure gradient existing across each structural hole is modelled by a *fluid138* element, extending from the fluid plane to the upper plate surface. The ambient relative pressure is imposed equal to zero in correspondence to external edges of the air gap and to the upper section of each hole; other constraints are used to impose the same pressure between the lower section of each hole and the air gap one. The simulation is repeated at different frequency values to detect $c$ and $k$ variation.

### 4.3. Modal projection method

This technique of numerical analysis allows the dynamic analysis of flexible oscillating structures; the assumption of a rigid motion is no more necessary and interactions between dynamic deformed shape and fluid pressure can be accounted for more accurately. The evaluation of damping and stiffness coefficients is based on the structure velocity profile determination, performed by a series of harmonic analysis.

By considering flexible structures, we observe a strong correlation between local velocity and local oscillation amplitude with the corresponding pressure in the fluid film. The modal projection technique considers the eigenvector (that contains all information about the modal deformed shape corresponding to a certain eigenfrequency) referred to a particular vibration mode (e.g. the first one) in the same way as an imposed velocity profile, which is applied on the structure. The subsequent gas pressure distribution is then evaluated for each fundamental frequency considered and, by knowing the structural geometry, the corresponding force is calculated; damping and stiffness parameters are obtained for each combination of structural modal deformed shape and imposed velocity profile.

The algorithm performed by the software can be described as follows [21]:
a. The fluid is excited by a velocity profile of an ideal structure that corresponds to the first eigenvector of the real structure (source mode); a harmonic response analysis is then performed to evaluate the pressure response in the desired frequency range.
b. Element pressure is integrated for each frequency to compute the element nodal force vector; both quantities are complex.

c. This force vector is multiplied for each eigenvector of the real structure (target modes), resulting the modal forces, indicators of how much the pressure distribution acts on each vibrating mode (back projection).
d. From real and imaginary part of modal forces damping and stiffness coefficients are extracted, corresponding to the first row of respecting matrices.
e. The procedure is repeated from point a. with the next eigenmode, till the last one considered.

When the computation procedure is complete, the damping and stiffness matrices are written as follows:

$$[K] = \begin{bmatrix} K_{1,1} & K_{1,2} & \dots & K_{1,n} \\ K_{2,1} & K_{2,2} & \dots & K_{2,n} \\ \dots & \dots & \dots & \dots \\ K_{n,1} & K_{n,2} & \dots & K_{n,n} \end{bmatrix} \qquad (12)$$

$$[C] = \begin{bmatrix} C_{1,1} & C_{1,2} & \dots & C_{1,n} \\ C_{2,1} & C_{2,2} & \dots & C_{2,n} \\ \dots & \dots & \dots & \dots \\ C_{n,1} & C_{n,2} & \dots & C_{n,n} \end{bmatrix} \qquad (13)$$

where $n$ is the number of modal shapes extracted. Terms on main diagonals represent effective stiffness and damping coefficients referred to the corresponding vibration mode, other terms represent coefficients connected to interaction effects (cross talk) between different modes, present in the case of non-uniform gap extension especially.

The structure is modelled by *solid45* elements and geometrical dimensions are derived from experimental measurements previously reported; the plate is fixed to four rectangular supports, laterally clamped. The fluid contained into air gap is modelled with the same elements used before and also pressure constraints are replicated.

## 5. EXPERIMENTAL MEASUREMENTS

The entire experimental FRF ($g$) is only considered in a small range around the resonance frequency, obtaining a discrete relation amplitude-frequency

$$a = g(f). \qquad (14)$$

This empirical curve is then interpolated by a polynomial function ($h$) of the six order, which numerical coefficients are numerically extracted by minimizing the quadratic errors function (17), where $n$ is the number of detected values.

$$a = h(f) \qquad (15)$$

$$h(f) = \alpha \cdot f^6 + \beta \cdot f^5 + \gamma \cdot f^4 + \qquad (16)$$
$$+ \delta \cdot f^3 + \varepsilon \cdot f^2 + \eta \cdot f + \vartheta$$

$$E(g,h) = \sum_{i=1}^{n} \left[ g(f)_i - h(f)_i \right]^2 \qquad (17)$$

When the analytical expression describing resonant peak is obtained, it is possible to extract its absolute maximum, corresponding to FRF maximum

$$|H|_{max} = max[h(f)], \qquad (18)$$

corresponding to the respective resonant frequency

$$f_{[h(f)]=max} = 2\pi\lambda_n. \qquad (19)$$

FRF value corresponding to the half-power level (characterized by 3dB decay from the peak) is then calculated as

$$[h(f)]_{3dB} = \frac{max[h(f)]}{\sqrt{2}} \qquad (20)$$

and two half-power points $f^I$ and $f^{II}$ are finally extracted

$$[H(f)]_{3dB} = H(f^I) \qquad (21)$$

$$[H(f)]_{3dB} = H(f^{II}). \qquad (22)$$

Half-power bandwidth is therefore defined as

$$\Delta\lambda = 2\pi f^{II} - 2\pi f^I = 2\zeta\lambda_n, \qquad (23)$$

allowing damping coefficient $\zeta$ estimation.

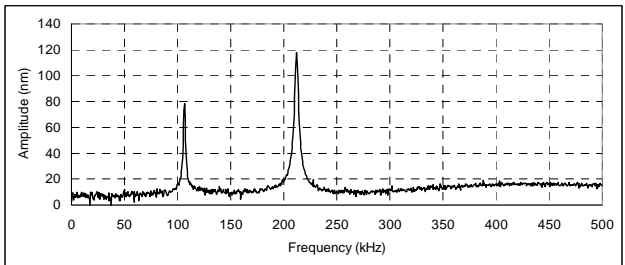

*Figure 4. FRF of structure type C.*

Figure 4 shows a typical interferometric FRF, referred to structure type C under 5V excitation amplitude and 2V offset, obtained by a frequency shift covering the range 0-500 kHz with 0.5 kHz steps.

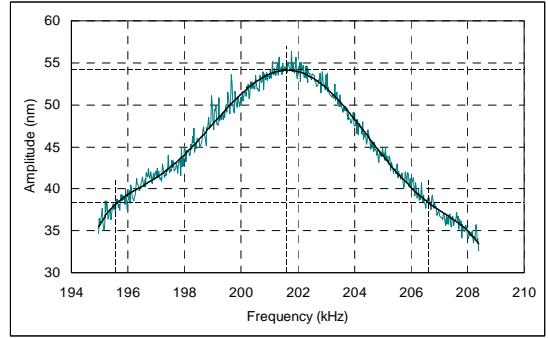

*Figure 5. Detection of resonant and half-power frequencies for modal damping and stiffness evaluation.*

Interpolation of experimental data with polynomial function is performed in correspondence of the resonant peak and characteristic frequencies are detected as represented in the figure 5. Next table 2 reports resonant and half-power frequency values obtained of each test structure.

| type | resonant frequency (kHz) | standard dev (kHz) | half-power frequencies (kHz) | |
|---|---|---|---|---|
| A | 201,637 | 5,418·10⁻² | 195,601 | 206,638 |
| B | 204,329 | 4,430·10⁻³ | 201,645 | 207,373 |
| C | 211,011 | 7,200·10⁻⁴ | 209,250 | 212,740 |
| D | 222,282 | 1,020·10⁻³ | 220,578 | 223,975 |
| E | 173,904 | 3,180·10⁻⁵ | 170,829 | 176,900 |
| F | 138,564 | 1,780·10⁻³ | 135,790 | 141,286 |

*Table 2. Experimental resonant frequency (statistically treated over 5 measurements) and half-power frequencies for each test structure type.*

## 6. RESULTS AND DISCUSSION

### 6.1. Imposed constant velocity method

Next figures, traced in a double-logarithmic scale, show their variation with respect to the actuation frequency; as expected, we observe an increasing of transferred stiffness from the fluid and a decreasing of damping.

By observing resonant frequency values reported in the table 2 for each test structure, we can realize that fluid damping transferred to the structure in correspondence of a resonant excitation is negligible, but the entity of transferred stiffness is considerably high.

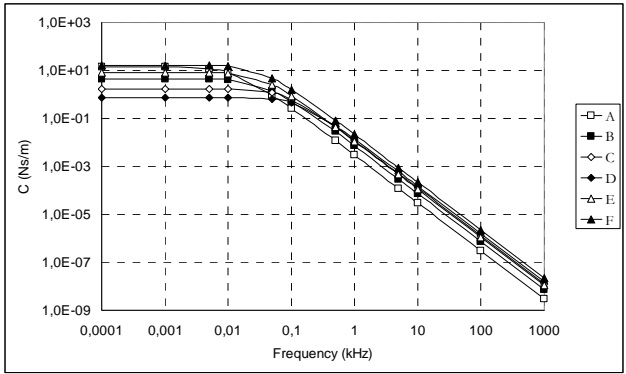

*Figure 6. Damping coefficient variation for each test structure computed by imposed velocity method.*

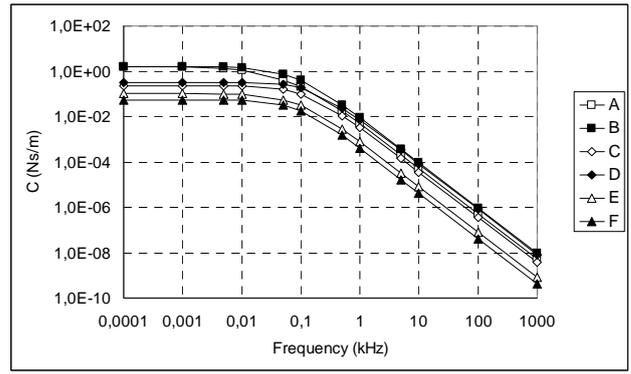

*Figure 9. Damping coefficient variation for each test structure computed by modal projection method.*

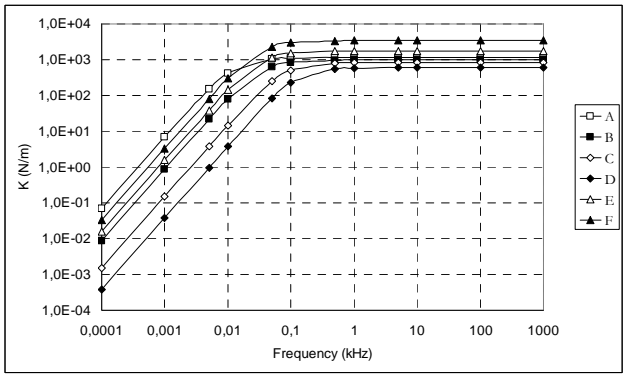

*Figure 7. Stiffness coefficient variation for each test structure computed by imposed velocity method.*

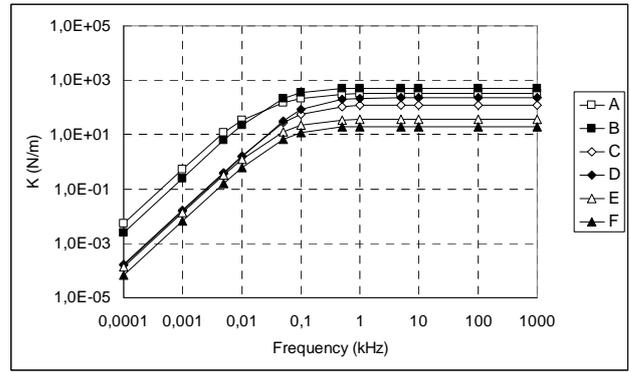

*Figure 10. Stiffness coefficient variation for each test structure computed by modal projection method.*

### 6.2. Modal projection method

A strong dependency of results from the finite element model meshing is observed; in particular an increase of the model elements number results in a progressive increase of both damping and stiffness values. This effect is represented in the figure 8, where C and K values are plotted with respect to the meshing element size that is expressed as a fraction of support width (*d*).

Figures 9-10, traced in a double-logarithmic scale, show coefficients variation with respect to the actuation frequency for each test structures computed with modal projection method.

### 6.3. Experimental measurements

Table 3 reports damping and stiffness coefficients experimentally evaluated for each test structure by the half-power method in resonating conditions.

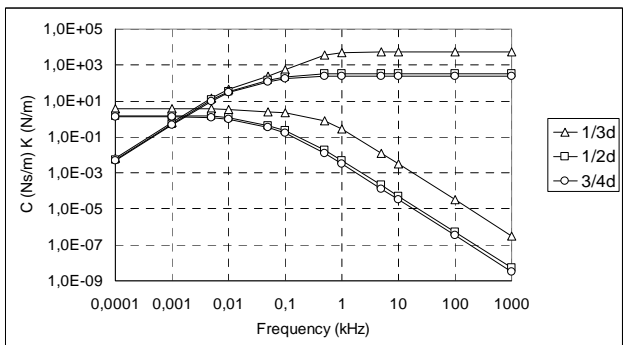

*Figure 8. Damping and stiffness variation with respect to meshing element size on test structure type A.*

| type | damping [Ns/m] | stiffness [N/m] |
|------|------------------|------------------|
| A | $4,738 \cdot 10^{-5}$ | $1,097 \cdot 10^{3}$ |
| B | $1,946 \cdot 10^{-5}$ | $8,912 \cdot 10^{2}$ |
| C | $9,863 \cdot 10^{-6}$ | $7,907 \cdot 10^{2}$ |
| D | $7,609 \cdot 10^{-6}$ | $6,954 \cdot 10^{2}$ |
| E | $3,822 \cdot 10^{-5}$ | $1,196 \cdot 10^{3}$ |
| F | $6,744 \cdot 10^{-5}$ | $1,480 \cdot 10^{3}$ |

*Table 3. Damping and stiffness resonant coefficients calculated by the experimental half-power method.*

The calculation is performed by modal mass value ($m_m$), in according to (24) and (25) equations:

$$c_m = 2m_m \zeta \lambda_n \qquad (24)$$

$$k_m = m_m \lambda_n^2 \qquad (25)$$

In the case of real systems is not possible to trace back the whole structure mass to a single degree of freedom because relevant inaccuracies may occur; however particular test structure geometry allows the consideration that a big part of the mass is involved in the dominant deformed shape, connected to a high participation factor associated to this eigenmode. This assumption derives from the comparison of the modal mass involved in the considered resonant deformation, calculated by a harmonic analysis, and the whole structure mass. Table 4 reports for each test structure the corresponding volume, the whole structure mass and the modal one; as expected mass ratios are near to the unity value. The critical damping $\zeta = c/c_{cr}$ directly extracted from half power calculations is also indicated.

| type | critical damping | volume [mm³] | modal mass [kg] | total mass [kg] | mass ratio |
|------|------|------|------|------|------|
| A | $2,737 \cdot 10^{-2}$ | $3,105 \cdot 10^5$ | $6,832 \cdot 10^{-10}$ | $7,442 \cdot 10^{-10}$ | 0,918 |
| B | $1,401 \cdot 10^{-2}$ | $2,653 \cdot 10^5$ | $5,407 \cdot 10^{-10}$ | $6,054 \cdot 10^{-10}$ | 0,893 |
| C | $8,270 \cdot 10^{-3}$ | $2,035 \cdot 10^5$ | $4,498 \cdot 10^{-10}$ | $5,080 \cdot 10^{-10}$ | 0,885 |
| D | $7,641 \cdot 10^{-3}$ | $1,693 \cdot 10^5$ | $3,565 \cdot 10^{-10}$ | $4,162 \cdot 10^{-10}$ | 0,857 |
| E | $1,745 \cdot 10^{-2}$ | $4,654 \cdot 10^5$ | $1,002 \cdot 10^{-9}$ | $1,059 \cdot 10^{-9}$ | 0,946 |
| F | $1,983 \cdot 10^{-2}$ | $8,982 \cdot 10^5$ | $1,953 \cdot 10^{-9}$ | $2,004 \cdot 10^{-9}$ | 0,974 |

*Table 4. Critical damping and masses comparison.*

### 6.4. Discussion

By analyzing numerical dynamic results at resonance we deduce an important stiffness increasing caused by the fluid presence, while the damping contribution is negligible. Result comparison must be consequently focused on stiffness results, while the smallness of damping values may produce estimation inaccuracies. Two finite element model results are in good agreement as the damping and stiffness variation with respect to the frequency testifies, also if the problem of meshing precision must be analyzed more accurately. Holes dimension effect is clearly visible in stiffness coefficient decreasing from test structure type A to D, that correspond to a hole section increasing, especially for constant velocity method simulations. Plate dimension effect is testified by an increase in stiffness coefficient with respect to the size, as test structure type E and F demonstrate.

Experimental measurements are conducted in ambient air, so corresponding results will take into account all dynamic interactions with surrounding fluid resulting in a structural and fluidic value of damping and stiffness.

Obtained values are however in good agreement with numerical ones.

Figures 11 and 12 reports a comparison of damping and stiffness coefficients variation with respect to the frequency derived by two different finite element model for test structure B and the same values experimentally obtained. Analogous graphics are obtained of each test structure type.

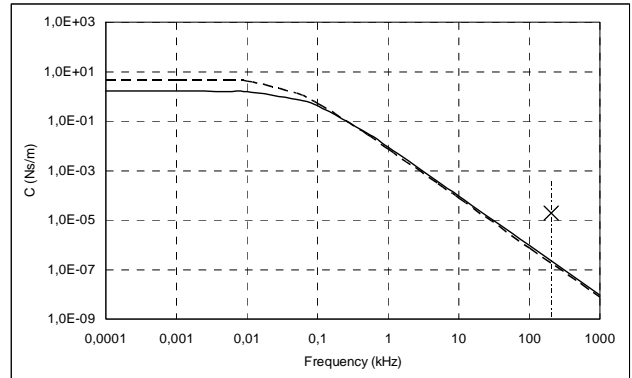

*Figure 11. Comparison between numerical and experimental damping coefficient for test structure B.*

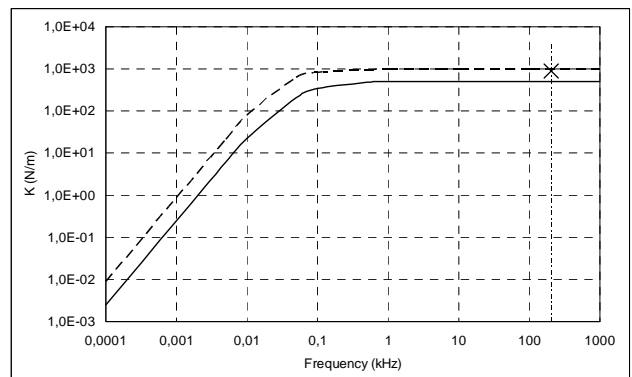

*Figure 12. Comparison between numerical and experimental stiffness coefficient for test structure B.*

## 7. CONCLUSIONS

Dynamic interactions between air present in the gap and oscillating plates with square holes are analyzed and discussed. Numerical models are realized to predict frequency response by the use of a finite element commercial software; two approaches are adopted: a constant imposed velocity method and a modal projection method.

Results are compared with experimental measurements and the effects of holes dimension and plate extension are discussed.

## 8. AKNOWLEDGEMENT

This work was partially funded by the Italian Ministry of University, under grant PRIN-2005/2005091729. Specimens were built by STMicroelectronics MEMS Business Unit (Cornaredo, Italy). Authors thank all the above involved institutions.